# Acceleration of magnetic dipoles by the sequence of current turns

*S. N. Dolya*

*Joint Institute for Nuclear Research, Joliot Curie str. 6, Dubna, Russia, 141980*

**Abstract**

Acceleration of magnetic dipoles is carried out by the running gradient of the magnetic field formed while sequent switching on the current turns. Magnetic dipoles, with a diameter of $d_{sh}$ = 60 mm and full length $l_{tot}$ = 1m, are pre-accelerated by using the gas-dynamic method to speed $V_{in}$ = 1 km / s, corresponding to the injection rate into the main accelerator. To prevent the turning of the dipoles by 180 degrees in the field of the accelerating pulse and focus them, the magnetic dipoles are accelerated inside the titanium tube. The magnetic dipoles have mass m = 2 kg and acquire the finite speed $V_{fin}$ = 5 km / s on the acceleration length $L_{acc}$ = 300 m.

**Introduction**

There is a known [1], method to accelerate the magnetic dipoles by the current pulse moving in the space. Using a large number of turns, in principle, allows one to reach a high finite speed of the magnetic dipoles. When the electric current is flowing in the coils, there is the magnetic field which pulls the magnetic dipole inside the coil with the current. After the magnetic dipole passing through the center of the coil, the magnetic field gradient changes its sign - due to this the magnetic dipole begins to accelerate in the opposite direction, i. e. inhibits. Therefore, to create a continuous acceleration, the current in the loop must quickly break off after the magnetic dipole passing through the coil center.

However, despite the apparent simplicity of this method, its practical use is accompanied with serious difficulties.

From the ferromagnetic materials being used for magnetic dipoles the most suitable is iron having a high specific magnetic moment and high Curie temperature. The specific magnetic moment is the property of the substance and cannot be increased. Furthermore, since the magnetic dipole must include a jet engine with the fuel supply and navigation devices, the specific magnetic dipole moment will be even less than that of the pure iron. That is why it is not possible to achieve a large finite speed of the magnetic dipole by using the acceleration method.

We explain the details. The magnetic moment per molecule in iron [2], page 524, is of the $n_b$ = 2.219 Bohr magneton. The table value of the Bohr magneton power [2], page 31, $m_b$ = 9.27 * $10^{-21}$ erg / Gs. Taking into account that the atomic weight of iron is: $A_{Fe}$ = 56, we find that the magnetic moment per nucleon in iron is: $m_{Fe} \approx 2 * 10^{-10}$ eV / (Gs*nucleon) and this value $m_{Fe}$ cannot be increased. Another principle disadvantage of this device is that the movement of the magnetic dipoles therein is unstable in the longitudinal direction. The reason is that at the attraction of the magnet poles with opposite signs there is phase instability [3].

Thus, if the magnetic dipole is slightly behind the running current pulse accelerating it, then it will turn out to be in a lower momentum field and, finally, will be forever behind it. If the magnetic dipole becomes too close to the current pulse, it will fall into a stronger field, being more and more attracted to the current pulse at the end. Finally, it will be ahead of it and turn by 180 degrees.

From the point of view of mutual positioning of the accelerating running pulse and the magnetic dipole, there is the only stable case when the pulse pushes (not pulls) the magnetic dipole. This means that the region of the phase (longitudinal) stability is located on the front slope of the traveling pulse. In the accelerator technology it is called the principle of phase stability [3].

The specific magnetic dipole moment can be, probably, increased (as compared with iron) by applying a current of the superconducting layer located inside the dipole. The magnetic field gradient, which accelerates the dipole, can be increased by superconductivity. The both ways lead to increasing the force accelerating the dipole: $F_z = m * dB_z / dz$, where m - specific magnetic dipole moment, $dB_z / dz$ - the magnetic field gradient.

We assume that the initial speed of the dipole is: $V_{sh}$ = 1 km / s and it is achieved by the gas-dynamic acceleration.

**1. Opportunity of increasing the specific magnetic moment in the magnetic dipole**

The specific magnetic dipole moment can be increased (as compared with iron), if to place the $Nb_3Sn$ superconducting winding inside the dipole and let the ring current flow through it.

Let us calculate how much the specific magnetic moment – the magnetic moment per unit of the mass of the magnetic dipole, will grow if to put a layer of superconducting Nb3Sn with radius $r_{cyl}$ = 3 cm and thickness $\delta_{cyl}$ = 0.2 cm, in its cylindrical part with a length equal to $l_{cyl}$=40 cm. We assume the current density in the superconductor to be equal [2], page 312, to $J_{ss}$ = 3 *$10^5$ A/$cm^2$. Then, linear density $j_{ss}$ current (A / cm) in such a superconducting layer is equal to: $j_{ss}$ (A / cm) = $J_{sp}$ * $\delta_{cyl}$ = 6 * $10^4$ A / cm. Such linear current density on the surface of the superconductor will create the magnetic field strength equal to $H_{ss}$ (kGs) = 1.226 * j (A / cm ) ≈ 70 kGs, that does not contradict the opportunity of achieving the current density of $J_{ss}$ ≈3 * $10^5$ A/$cm^2$ [2], page 312. The total current flowing in the superconducting layer is equal to $I_{SC}$ = $j_{ss}$ *$l_{cyl}$ = 2.4 * $10^6$ A, it will lead to the appearance of magnetic moment **$M_{ss}$** = $I_{SC}$ * $\pi r_{cyl}^2$ = 6.8 * $10^7$ A * $cm^2$ or, in the CGS system, **$M_{ss}$** = 6.8 * $10^6$ erg / Gs.

The total mass of the superconducting layer can be calculated from the following: the density of $Nb_3Sn$ superconductor is $\rho_{Nb3Sn}$ = 8 g/$cm^3$, the atomic mass: A = 400 and the total volume of superconductor $V_{sp}$ = 150 $cm^3$ contains $N_{Nb3Sn}$ = 7.2 *$10^{26}$ nucleons. The specific magnetic moment, the magnetic moment per unit of the mass (nucleon), is then equal to: **M**ss = $m_{ss}$ /$N_{Nb3Sn}$ = =0.94 * $10^{-20}$ erg / (Gs *nucleon) = 5.9 * $10^{-9}$ eV / (Gs * nucleon), that is approximately 30 times greater than in iron [2], page 524.

**2. Ways to achieve the required parameters of acceleration**

Let the mass of the superconductor in the magnetic dipole be $m_{Nb3Sn}$ = 1.2 kg, the mass of the jet engine, fuel, navigation control devices is equal $m_{Fuel}$ = 0.8 kg, then the specific magnetic moment in the magnetic dipole will be equal to: **$m_0$** = 3.5 * $10^{-9}$ eV / (Gs *nucleon), that is approximately 17 times greater than in iron. The pulse duration of the magnetic field can be determined from the following considerations. To place the magnetic dipole on the length of the pulse accelerating it, slowdown wavelength must be of the order of: $\lambda_{slow}$ = 4 m. The time period $T_0$ of the corresponding wave is determined from the following relationship:

$$\lambda_{slow} = V_{sh} *T_0, \qquad (1)$$

where we find that $T_0$ = 4 ms and the wave frequency corresponding to this period is equal to: $f_0$ = 250 Hz.

## 3. Selecting the thickness of the barrel wall of the Gauss gun

The region of the phase stability in azimuthally symmetric wave corresponds to the region of the radial instability. The magnetic dipole will push itself from the pole of the same sign, but, first of all, it will seek to turn around by $180^0$ and be attracted by the opposite sign poles. To prevent the radial escape of the dipole and its reversal by $180^0$ in the field of the pulse accelerating is possible if to place the dipole inside the titanium tube whose inner diameter is the same as the outer diameter of the dipole. The titanium tube wall thickness must be of such value to let the external magnetic field freely without distortion penetrate inside. It means that it should be much smaller than the skin-layer depth in titanium.

Electrical resistance of copper $\rho_{Cu} = 1.67 *10^{-6}$ Ohm * cm, titanium $\rho_{Ti} = 55 *10^{-6}$ Ohm *cm, [2], page 305, the conductivity $\sigma$ (dimension $\sigma$ is 1 / s) are related with a specific resistance value: $\sigma = 9 * 10^{11} / \rho$ for copper the conductivity is : $\sigma_{Cu} = 5.4 * 10^{17}$ 1 / s, for titanium $\sigma_{Ti} = 3.23 * 10^{16}$ 1 / s. This allows one to calculate the depth of the skin-layer and, thereby, to calculate the possible thickness of the tube wall, where the magnetic dipole will be accelerated.

Let us find the thickness of the skin - layer for titanium for frequency $f_0 = 250$ Hz. It can be calculated by the following formula:

$$\delta_{Ti} = c/2\pi \, (f_0 \sigma_{Ti})^{1/2} = 1.68 \text{ cm.} \qquad (2)$$

This means that the wall thickness of the titanium tube $\Delta h_{Ti}$, wherein the magnetic dipole has to be accelerated, can be chosen to be equal to: $\Delta h_{Ti} = 2$ mm.

## 4. Interaction of the dipole with the magnetic field gradient

For stable acceleration of magnetic dipoles it is necessary to "switch on" consequently the magnetic coils according to the dipole move. The magnetic field of the coil with a current can be written as follows:

$$B_z = I_0 * r_0^2 / [\, 2 * (r_0^2 + z^2)^{3/2}], \qquad (3)$$

where: $I_0$ - current in the loop, Ampere, $r_0$ - the radius of the coil with a current, cm, z - the distance from the coil plane to the observation point.

In comparison of the multi section Gauss gun [1] the corresponding coil here is necessary "to be switched on" after the magnetic dipole passage through the coil center but not to join the coil switched on in advance.

*4.1. Acceleration of the magnetic dipole by the current single coil*

We differentiate expression (3) with respect to z and obtain a formula for the magnetic field gradient:

$$dB_z / dz = (3/2) * r_0^2 * I_0 * z / (r_0^2 + z^2)^{5/2}. \qquad (4)$$

From this formula it is clear that the gradient field is zero in the coil plane at z = 0.

We assume that at a distance of the order of the radius of the turn, the speed of the dipole varies slightly, i.e. it is possible to change variable z for $V_{sh}t$. The specific magnetic dipole moment increases while the dipole passing through the center of the coil according to the law:

$$\mathbf{m} = 2\mathbf{m_0} * z / l_{sh}, \qquad (5)$$

where $\mathbf{m_0} = 3.5 * 10^{-9}$ eV / (Gs * nucleon), $l_{sh}$ - length of the dipole. The current in the coil after switching increases linearly with time, according to the law:

$$I \approx I_0 * (4t/T_0), \qquad (6)$$

where $T_0$ - time period of the slowdown wavelength.

The force influencing the dipole from the coil side is:

$$F_z = \mathbf{m_0} * (z / l_{sh}) * (3/2) * r_0^2 * z * I_0 * 4 (t/T_0) / (r_0^2 + z^2)^{5/2}. \qquad (7)$$

Substituting t for $z / V_{sh}$ and integrating over z, we obtain an expression for the energy gain rate while passing one loop by the magnetic dipole:

$$\Delta W = \int F_z dz =$$
$$= \int \mathbf{m_0} * (z / l_{sh}) * (3/2) * r_0^2 * z * I_0 * 4 (z/V_{sh}T_0) /(r_0^2 + z^2)^{5/2} dz, \qquad (8)$$

or

$$\Delta W_1 = 12\mathbf{m_0} * (r_0^2/l_{sh}) *(I_0/V_{sh}T_0) \int_0^{l_{sh}/2} [z^3 / (r_0^2 + z^2)^{5/2}] \, dz. \qquad (9)$$

Integration in (9) should be performed till the distance approximately equal to half of the length of the magnetic dipole: $l_{sh} = 1m$. The same order should be the turn radius: $r_0 = 1m$. After the magnetic dipole passing a distance $l_{sh} / 2$, its magnetic moment does not any longer increase and the dipole magnet will be just repelled by a coil with a current.

The corresponding set of energy can then be written as follows:

$$\Delta W_2 = 6\mathbf{m_0} * r_0^2 *(I_0/V_{sh}T_0) \int_{l_{sh}/2}^{r_0} [z_2 / (r_0^2 + z^2)^{5/2}] \, dz. \qquad (10)$$

Substituting numerical values of $r_0 = 1m$, $l_{sh} = 1m$, calculating the integrals and summing $\Delta W_1$ and $\Delta W_2$, we find that the energy acquired during the passage of one current loop by the magnetic dipole is equal to:

$$\Delta W \approx 12\mathbf{m_0} * (r_0^2/l_{sh}) * (I_0/V_{sh}T_0) * 5.5 * 10^{-2}. \qquad (11)$$

*4.2. Acceleration of the magnetic dipole by consequence of current turns*

We assume that the consequence of current turns is as follows: per 1m there are $10^3$ turns ($10^3 / m$), the current in each coil is assumed to be equal to: $I_0 = 150$ kA. Assuming $\mathbf{m_0} = 3.5 * 10^{-9}$ eV / (Gs *nucleon) and averaging the action of the turns on the magnetic dipole with a coefficient of ½, we finally obtain the formula for the energy gain rate of the magnetic dipole:

$$\Delta W = 4.33 * 10^{-4} \text{ (eV / nucleon * m)}. \qquad (12)$$

Multiplying $\Delta W = 4.33 * 10\text{-}4$ (eV / nucleon * m) by the length of the acceleration $L_{acc} = 300$ m, we find the finite energy of the magnetic dipoles: $W_{fin} = 0.13$ eV / nucleon, that corresponds to the finite speed of the magnetic dipoles $V_{fin} = 5$ km / s.

Figure 1 shows a diagram of the device.

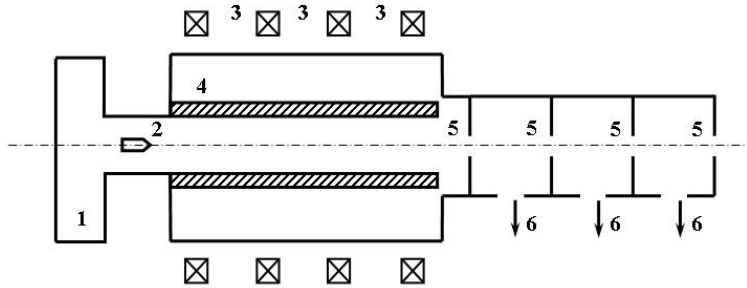

Fig.1. (1) - gun, (2) - magnetic dipoles, (3) - current coils,
(4) - titanium tube, (5) - pulse diaphragm, (6) - pumping.

**Conclusion**

While increasing the diameter of the magnetic dipole its total magnetic moment grows as the area of its turn, i.e. proportionally, as $r^2$. The dipole mass, at a constant current density increases as the perimeter of the loop, i.e., linearly with the radius. Thus, the specific magnetic dipole moment will grow linearly with increasing of the turn radius.

Literature

1. http://ru.wikipedia.org/wiki/Пушка_Гаусса

2. Tables of physical quantities. Reference ed. Kikoin,
   Moscow, Atomizdat, 1976

3. V. I. Veksler, Dokl. USSR Academy of Sciences, 1944, v. 43,
   Issue 8, p. 346, E. M. McMillan, Phys. Rev. 1945, v. 68, p. 143